\documentclass[%
preprint,
showpacs,preprintnumbers,
 amsmath,amssymb,
 aps,
prstab,
floatfix,
]{revtex4-1} 
\pdfoutput=1 
\usepackage{graphicx}
\usepackage{dcolumn}
\usepackage{bm}
\usepackage{hyperref}
\usepackage{booktabs}
\usepackage{soul}

\newcommand{\cv}[1]{c_\mathrm{V}^{#1}}
\newcommand{\cvhat}[1]{\hat c_\mathrm{V}^{#1}}
\newcommand{\zv}[1]{Z_\mathrm{V}^{#1}}
\newcommand{\meas}[1]{\int\mathrm d#1\,}
\newcommand{\eval}[1]{\langle#1\rangle}
\newcommand{\order}[1]{\mathrm{O}(#1)}
\newcommand{\SU}[1]{\mathrm{SU}(#1)}

\begin{document}

\preprint{MITP, HIM}

\title{Non-perturbative improvement of the vector current in Wilson lattice QCD}

\author{Tim Harris$^a$}
 \email{harris@kph.uni-mainz.de}
\author{Harvey B.\ Meyer$^{a,b}$}%
 \email{meyerh@kph.uni-mainz.de}
\affiliation{$^a$Helmholtz~Institut~Mainz, D-55099 Mainz, Germany \\ 
$^b$PRISMA Cluster of Excellence and Institut f\"ur Kernphysik 
Johannes Gutenberg-Universit\"at Mainz, D-55099 Mainz, Germany }

\date{\today}

\begin{abstract}
Many observables of interest in lattice QCD are extracted from correlation functions
involving the vector current. If Wilson fermions are used, it is therefore 
of practical importance that, besides the action, the current be O($a$) improved
in order to remove the leading discretization errors from the observables.
Here we introduce and apply a new method to determine the improvement coefficient
for the two most widely used discretizations of the current.
\end{abstract}

\maketitle


\section{\label{sec:introduction}Introduction}
Lattice QCD is a powerful tool to calculate the predictions of Quantum
Chromodynamics in its non-perturbative regime.  While the quantum
field theory is regularized by discretizing it on a lattice,
ultimately the quantities of interest -- for instance, ratios of
hadron masses -- must be determined in the limit where the cutoff is
removed. For numerical purposes, it is computationally advantageous to
accelerate the approach to the continuum by removing the leading
cutoff effects. In particular, Symanzik's continuum effective
theory~\cite{Symanzik:1983dc,Symanzik:1983gh} can be used to remove
the $\order{a}$-cutoff effects which appear generically when using the
Wilson fermion action in lattice QCD simulations~\cite{Wilson:1974sk}.
To eliminate $\order{a}$-cutoff effects in the hadronic spectrum it
suffices to improve the action by introducing the dimension-five
Sheikholeslami-Wohlert term~\cite{Sheikholeslami:1985ij} with a non-perturbatively determined
coefficient $c_\mathrm{sw}$~\cite{Luscher:1996sc}.  However,
the addition of higher-dimensional counterterms to local operators is
also necessary for the improvement of their matrix elements, along
with the determination of the corresponding improvement coefficients.


In the following, we focus on the vector current, which requires a single
O($a$)-improvement term.
Estimates of the relative contribution of the improvement term
evaluated with the perturbative improvement coefficient may suggest
that the effect of the improvement in correlation functions would be
small for the local vector current~\cite{Guagnelli:1997db}.  However,
an improvement condition based on chiral Ward identities previously
used to determine the improvement coefficient $\cv{l}$ non-perturbatively in pure gauge theory
with~\cite{Guagnelli:1997db} and without~\cite{Bhattacharya:1999uq}
Schr\"odinger functional boundary conditions indicated significant
deviations from the tree-level result.

In this work, we describe a simple prescription for the non-perturbative
determination of the improvement coefficients, $\cv{l}$ and $\cv{c}$,
for the local and conserved (point-split) isovector vector currents,
defined below.  In the following section we report large differences
between the lowest-order perturbative estimates and our
non-perturbative evaluation of the improvement coefficients with
$N_\mathrm{f}=2$ Wilson clover fermions.  In
section~\ref{sec:continuum_limit} we demonstrate the effects of the
improvement on the scaling of an observable in the continuum limit.

\section{Theory background and a new improvement condition}

We use the O($a$)-improved Wilson fermion action with the non-perturbatively tuned value of $c_{\rm sw}$~\cite{Jansen:1998mx}.
The two discretizations of the continuum vector current that we employ are
\begin{align}
    \label{eq:local_current}
    (V)^l_\mu(x) &= \overline{\psi}(x) \gamma_\mu \frac{\tau_3}{2} \psi(x), \\
    \label{eq:conserved_current}
    (V)^c_\mu(x) &= \frac{1}{2}\left( 
        \overline{\psi}(x + a\hat\mu) (1 + \gamma_\mu) U_\mu^\dagger(x)\frac{\tau_3}{2} \psi(x)\right. 
        \nonumber\\
        &\left.\qquad\qquad- \overline{\psi}(x) (1 - \gamma_\mu) U_\mu(x)\frac{\tau_3}{2} \psi(x)
        \right).
\end{align}
The renormalized improved current for $i=l,c$ is defined by~\cite{Luscher:1996sc}
\begin{align}
    \label{eq:renormalized_current}
    (V_\mathrm{R})^i_\mu(x) &= \zv{i} ( 1 + b^i_\mathrm{V} am_\mathrm{q}) (V_\mathrm{I})^i_\mu(x), \\
    \label{eq:improved_current}
    (V_\mathrm{I})^i_\mu(x) &= (V)^i_\mu(x) + a\cv{i} \partial_\mu T_{\mu\nu}(x),
\end{align}
where the lattice discretization of $\partial_\mu T_{\mu\nu}(x)$ will
be discussed later and $\zv{l}\equiv\zv{}$ in the notation of~\cite{Luscher:1996sc}, 
while $\zv{c}=1$.  
The on-shell improvement of the vector current is required in many lattice
studies: hadronic form factors, the hadronic contributions to $(g-2)_\mu$ 
and thermal correlation functions related to the dilepton production rate, 
to mention a few.

While the local vector current requires improvement only at one-loop
order~\cite{Sint:1997jx} the conserved vector current requires
improvement at tree level also in the massless limit,
\begin{align}
    \label{eq:perturbative_cvl}
    \cv{l} &= -0.01225(1) \times C_\mathrm{F} \times g_0^2 + \order{g_0^4},\\
    \label{eq:perturbative_cvc}
    \cv{c} &= \frac{1}{2} + \order{g_0^2},
\end{align}
where $C_F= (N^2-1)/(2N)$ is the quadratic Casimir in the fundamental
representation for gauge group $\SU{N}$.

\subsection{\label{sec:improvement_condition}Improvement condition for $\cv{l,c}$}

\begin{figure*}[t]
    \centerline{
    \includegraphics[scale=0.8]{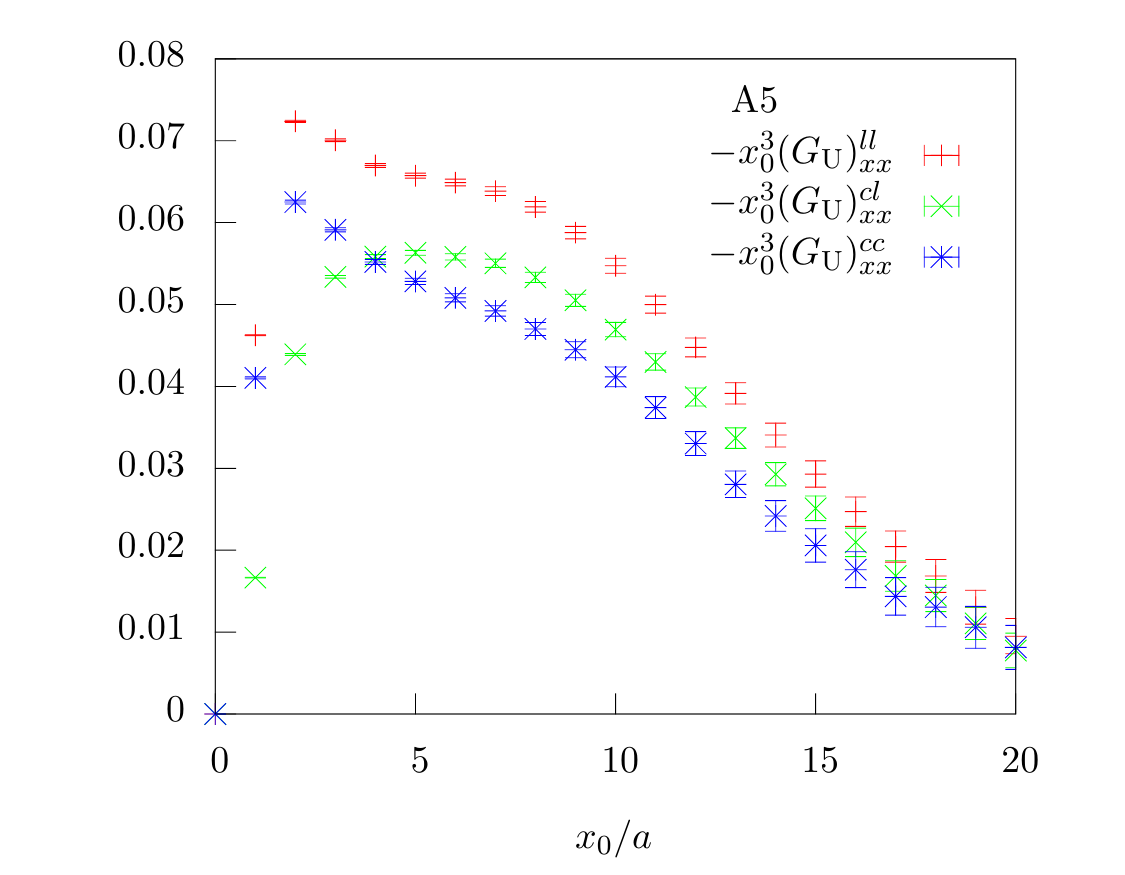}%
    \includegraphics[scale=0.8]{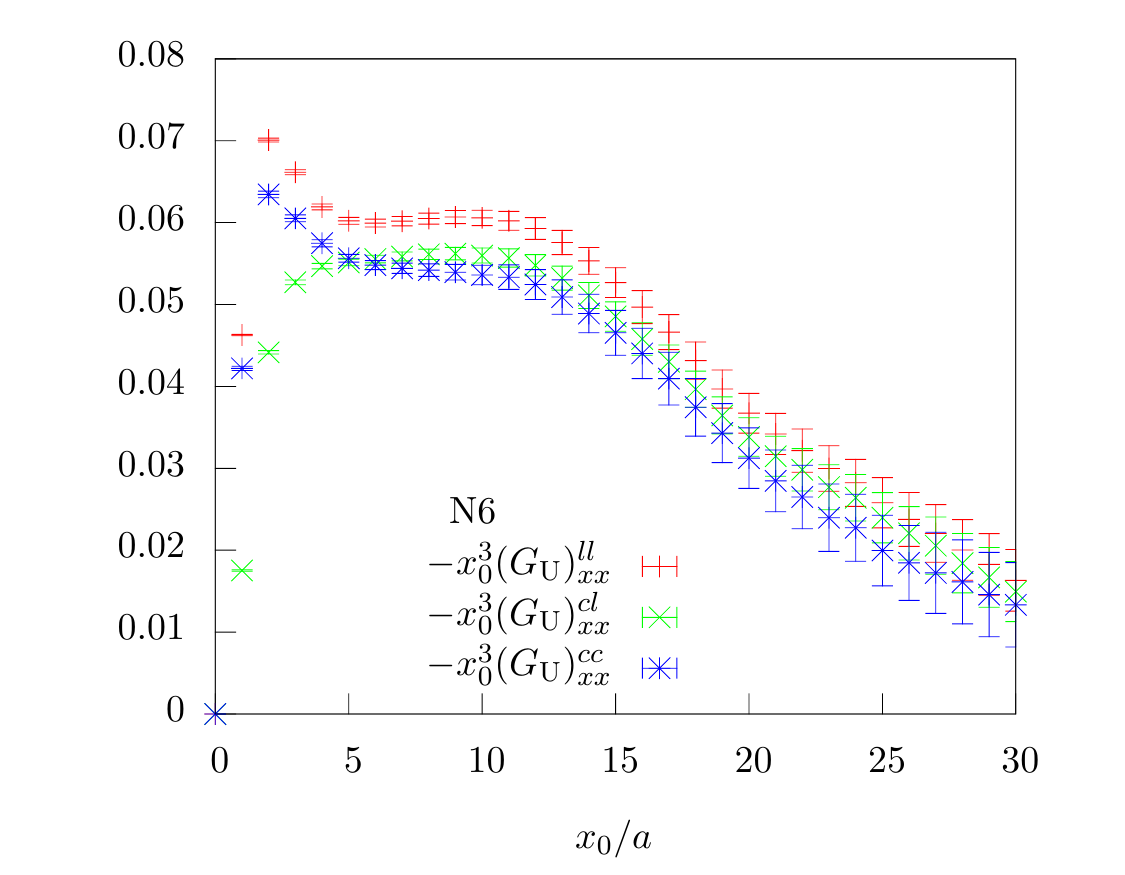}
    }
    \caption{Unimproved correlators for ensemble A5 (left) and N6 (right).}
    \label{fig:unimproved_correlators}
\end{figure*}
The proposed improvement condition is based on two discretizations of
the vector current defined in eq.~(\ref{eq:local_current}) ($l$) and
eq.~(\ref{eq:conserved_current}) ($c$).  
The main observable we consider is the vector current correlator
\begin{align}
    \label{eq:vector_correlator}
    \nonumber
    (G_\mathrm{I})^{ij}_{\mu\nu}(x_0)&\equiv(G_\mathrm{U})^{ij}_{\mu\nu}(x_0)\\
        \nonumber
    &+ \zv{i}\zv{j}\meas{^3x}
         \Big\langle a\cv{j}(V)_\mu^i(x_0,\bm x)\partial_\rho T_{\nu\rho}(0) \\
        & + a\cv{i}\partial_\rho T_{\mu\rho}(x_0,\bm x)(V)_\nu^j(0) \Big\rangle,
\end{align}
with
\begin{align}
    \label{eq:unimp_vector_correlator}
    (G_\mathrm{U})^{ij}_{\mu\nu}(x_0)\equiv
            \zv{i}\zv{j}\meas{^3x}
             \Big\langle (V)_\mu^i(x_0,\bm x)(V)_\nu^{j}(0) \Big\rangle,
\end{align}
where  we have indicated its dependence on the discretization of the current at finite lattice spacing,
 $ij=ll,cl,cc$. We will use periodic boundary conditions in space and thermal boundary conditions  
in time, though our method is more generally applicable, for instance to open boundary conditions in time.
In ref.~\cite{Francis:2013jfa}, two discretizations of the unimproved vector current correlator
demonstrated significant differences in the region of $t \lessapprox
0.5$fm at intermediate lattice spacings corresponding to bare lattice
couplings $\beta=5.3$ with $N_\mathrm{f}=2$ Wilson clover fermions.
This suggests that the three independent discretizations of the
temporal vector current correlator could be used to formulate an improvement condition. 

In figure~\ref{fig:unimproved_correlators}, we illustrate the
discrepancies between the three discretizations for two lattice
spacings corresponding to bare lattice couplings of $\beta=5.2$ and
$\beta=5.5$ in the left and right panels respectively.  The details of
the ensembles and the number of measurements are listed in
table~\ref{tab:cls_nf2}.  The non-perturbative renormalization
constant, $Z_\mathrm{V}$, is taken from ref.~\cite{DellaMorte:2005rd}.
{Note that a more precise non-perturbative result for $Z_\mathrm{V}$ has been reported in ref.~}\cite{Brida:2014zwa}.
As expected, the differences between the two discretizations are
reduced as the lattice spacing decreases.

Demanding that three discretizations of the temporal vector current correlator agree at a particular
Euclidean time $x_0$,
\begin{align}
    \label{eq:improvement_condition}
    (G_\mathrm{I})^{ll}_{xx}(x_0) \stackrel{!}{=}
        (G_\mathrm{I})^{cl}_{xx}(x_0) \stackrel{!}{=}
        (G_\mathrm{I})^{cc}_{xx}(x_0),
\end{align}
allows one to solve the following $2\times2$ linear system for the improvement coefficients $\cv{l,c}(x_0)$:
\begin{align}
    \label{eq:linear_system}
    \nonumber
    &\left(
    \begin{array}{cc}
        2\zv{}G^{lT}_{xx}(x_0) - G^{cT}_{xx}(x_0) & -G^{lT}_{xx}(x_0)\\
        -\zv{}G^{cT}_{xx}(x_0) & 2G^{cT}_{xx}(x_0) - \zv{}G^{lT}_{xx}(x_0) \\
    \end{array}
    \right) \\
    &\quad\times
    \left(
    \begin{array}{c}
        \cv{l}\\
        \cv{c}
    \end{array}
    \right)
    = \frac{1}{a}
    \left(
    \begin{array}{c}
        (G_\mathrm{U})_{xx}^{cl}(x_0) - (G_\mathrm{U})_{xx}^{ll}(x_0) \\
        (G_\mathrm{U})_{xx}^{cl}(x_0) - (G_\mathrm{U})_{xx}^{cc}(x_0) 
    \end{array}
    \right),
\end{align}
where 
\begin{align}
    \label{eq:vector_tensor_correlator}
    G^{iT}_{\mu\nu}(x_0)=\meas{^3x}\eval{(V)^i_\mu(x_0,\bm
    x)\partial_\rho T_{\nu\rho}(0)}.
\end{align}
%
{Translation invariance and time-reversal antisymmetry of the vector-tensor correlation function was used to simplify the system}~(\ref{eq:linear_system}).
{The correlators appearing in }(\ref{eq:linear_system}) {may be averaged over the spatial components to improve the signal.}
We define the improvement coefficient to be
$\cvhat{l,c}=\cv{l,c}(x_0)$ for some choice of $x_0$.  The method is viable in practice 
provided that a signal exists both for the r.h.s. and the
determinant of the linear operator on the l.h.s. of
eq.~(\ref{eq:improvement_condition}).  The results for $\cv{}$ obtained
using different legitimate prescriptions will in general differ by
O($a$) corrections.  Ideally, one would choose $x_0$ in a region where
there is both a signal and higher-order lattice artifacts are highly
suppressed.  This improvement condition can be implemented directly in
a finite volume and is straightforward to compute, not requiring the
three-point functions of ref.~\cite{Guagnelli:1997db}.  Although the
quark mass-dependence of the renormalization factor is neglected in
this improvement condition, {namely the $b_\mathrm{V}$ term in }eq.~(\ref{eq:renormalized_current}), due to the smallness of the quark mass
these effects ought to be small.  Furthermore, in the following
section numerical evidence demonstrates that no dependence on the
quark mass is likely to be observed in these improvement coefficients.

\subsubsection{\label{sec:scheme}Discretization of $\partial_\mu$ and $T_{\mu\nu}$}

In the improvement term, we use the local discretization of the tensor current,
\begin{align}
    T_{\mu\nu}(x) = -\frac{1}{2}\bar\psi(x) [\gamma_\mu,\gamma_\nu] \frac{\tau_3}{2}\psi(x),
\end{align}
with the same spacetime argument as the vector current.
The choice of the discretization of $\partial_\mu$ affects only
higher-order lattice artifacts, which nevertheless can be large.  In
ref.~\cite{Gockeler:2003cw}, the improvement of the conserved current (\ref{eq:conserved_current}) was considered.
The effect of using the symmetric derivative $\tilde\partial_\nu$ and the tensor current 
averaged over sites $x$ and $(x+a\hat\mu)$
was examined at one-loop level in lattice perturbation
theory.  While with this choice the identity $\partial^{*}_\mu
(V_\mathrm{R})^c_\mu=0$ still holds in on-shell correlation functions,
it was noted to introduce large higher-order lattice artifacts to the
connected part of the hadronic vacuum polarization tensor.  Therefore,
for the tensor current
 with time argument $x_0$ (assumed positive) 
 in correlation function (\ref{eq:vector_correlator}), 
we choose the forward finite-difference derivative, while 
for the tensor current at the origin we used the backward derivative.
We remark that it is admissible to use different discretizations of
$\partial_\mu$ in the determination of the improvement coefficient and
subsequently in matrix elements of the improved operator~\footnote{
It would however not be possible to use
 a point-split tensor current for the improved operator in conjunction
 with the improvement coefficient determined from the local one.  These
 definitions differ already at $\order{g_0^2}$.}.

\begin{table}
    \centering
    \begin{ruledtabular}
        \begin{tabular}{ccccccc}
            $\beta$    & $a$(fm)     & $m_\pi$(MeV)       & $T\times L^3$     & \# cfgs        & \# sources & name \\
            5.2 & 0.079 & 312 & $64\times32^3$ & 217 & 8   & A5a \\
                &       &     &                & 404 & 4   & A5d \\
            5.3 & 0.063 & 451 & $64\times32^3$ & 421 & 4   & E5g \\
                &       & 324 & $96\times48^3$ & 294 & 4   & F6 \\
            5.5 & 0.050 & 340 & $96\times48^3$ & 568 & 1   & N6 \\
        \end{tabular}
    \end{ruledtabular}
    \caption{Details of CLS $N_\mathrm{f}=2$ ensembles and number of measurements used in this work.}
    \label{tab:cls_nf2}
\end{table}

\begin{figure*}[t]
    \centerline{
    \includegraphics[scale=0.8]{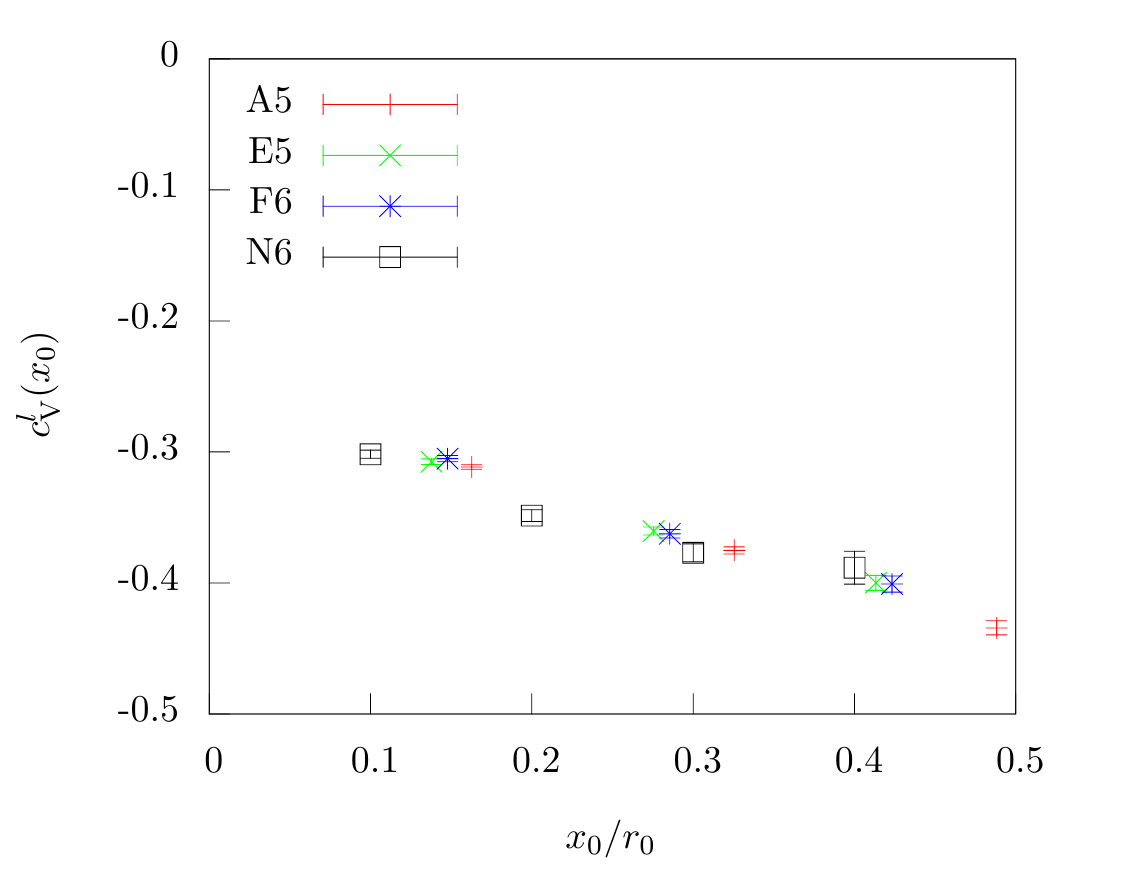}%
    \includegraphics[scale=0.8]{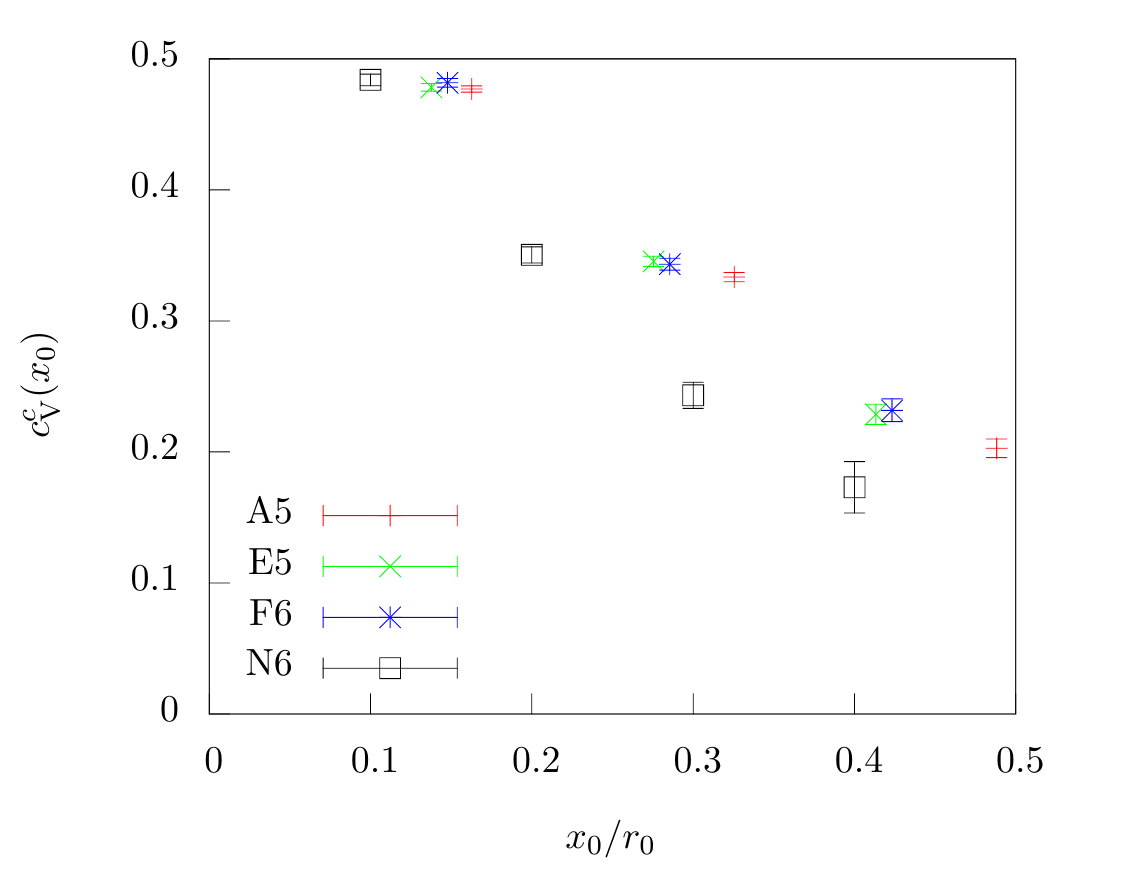}
    }
    \caption{Non-perturbatively determined improvement coefficient $\cv{l}$ (left) and $\cv{c}$ (right).
        The blue bursts have been displaced horizontally for clarity.}
    \label{fig:cv}
\end{figure*}

\section{Evaluation of $\cvhat{c,l}$}

Using the improvement condition eq.~(\ref{eq:improvement_condition})
we determined $\cv{l,c}(x_0)$ at three lattice spacings on ensembles
with $N_\mathrm{f}=2$ Wilson clover fermions {with non-perturbatively tuned value of $c_\mathrm{sw}$}~\cite{Jansen:1998mx} and the plaquette action.
The most relevant parameters are given in table \ref{tab:cls_nf2}; for
more details on the ensembles, see~\cite{Fritzsch:2012wq}, where the
action is also given explicitly.  In figure~\ref{fig:cv} we show the
dependence of $\cv{l,c}$ on the choice of $x_0$ for ensembles $\mathrm
A5$, $\mathrm F6$ and $\mathrm N6$ with bare lattice couplings
$\beta=5.2,5.3$ and $5.5$, respectively, and almost identical pion
masses.  Additionally, we compare with another ensemble, $\mathrm E5$,
with $\beta=5.3$ and a larger quark mass.  In figure~\ref{fig:cv}
(left), some evidence can be seen for a plateau in the value of
$\cv{l}$ at the smallest lattice spacing corresponding to $\beta=5.5$
(black squares).  For $\cv{l}$, no significant changes can be observed
as a function of the lattice spacing.  The right-hand panel of
figure~\ref{fig:cv} shows there is a greater dependence on the cutoff
for $\cv{c}$.  We are unable to distinguish any dependence on the
quark mass.

Our choice for the improvement coefficients is
$\cvhat{l,c}=\cv{l,c}(x_0/a=3)$ which is used in the rest of this
work.  Our results are given in table \ref{tab:results}. 

\begin{table}
    \centering
    \begin{ruledtabular}
        \begin{tabular}{cdd}
            Ensemble & \multicolumn{1}{c}{$\cvhat{l}$}     & \multicolumn{1}{c}{$\cvhat{c}$}       \\
            A5       & -0.434(5)    &  0.203(7)       \\
            E5       & -0.400(6)    &  0.229(8)       \\
            F6       & -0.401(6)    &  0.232(9)       \\
            N6       & -0.377(7)    &  0.243(10)
        \end{tabular}
    \end{ruledtabular}
    \caption{Results for improvement coefficients for the local and conserved vector currents for the ensembles listed in table~\ref{tab:cls_nf2}.
    The statistical error is estimated by single-elimination jackknife resampling.}
    \label{tab:results}
\end{table}
This choice leads to values of $\cvhat{l}$ which deviate significantly from the
perturbative estimate of eq.~(\ref{eq:perturbative_cvl}).  We can make
a rough comparison with the value of $\cv{l}$ determined in the
quenched theory~\cite{Guagnelli:1997db} based on the observation that,
at fixed bare coupling, the improvement coefficients are independent of $N_\mathrm{f}$ to
one-loop in perturbation theory. We note that the value we obtain at $\beta=5.3$
is quite similar to the value obtained at $\beta=6.0$ in the quenched theory.

\begin{figure*}[t]
    \centerline{
    \includegraphics[scale=0.8]{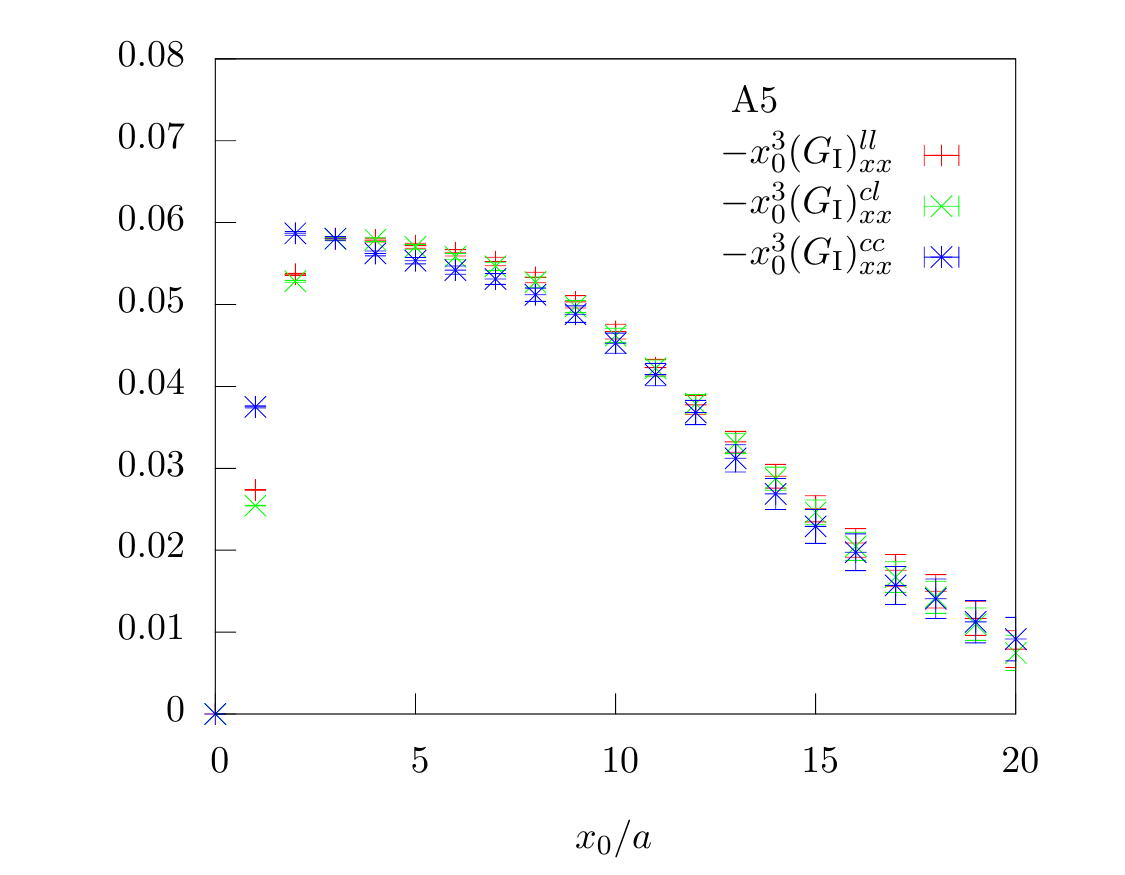}%
    \includegraphics[scale=0.8]{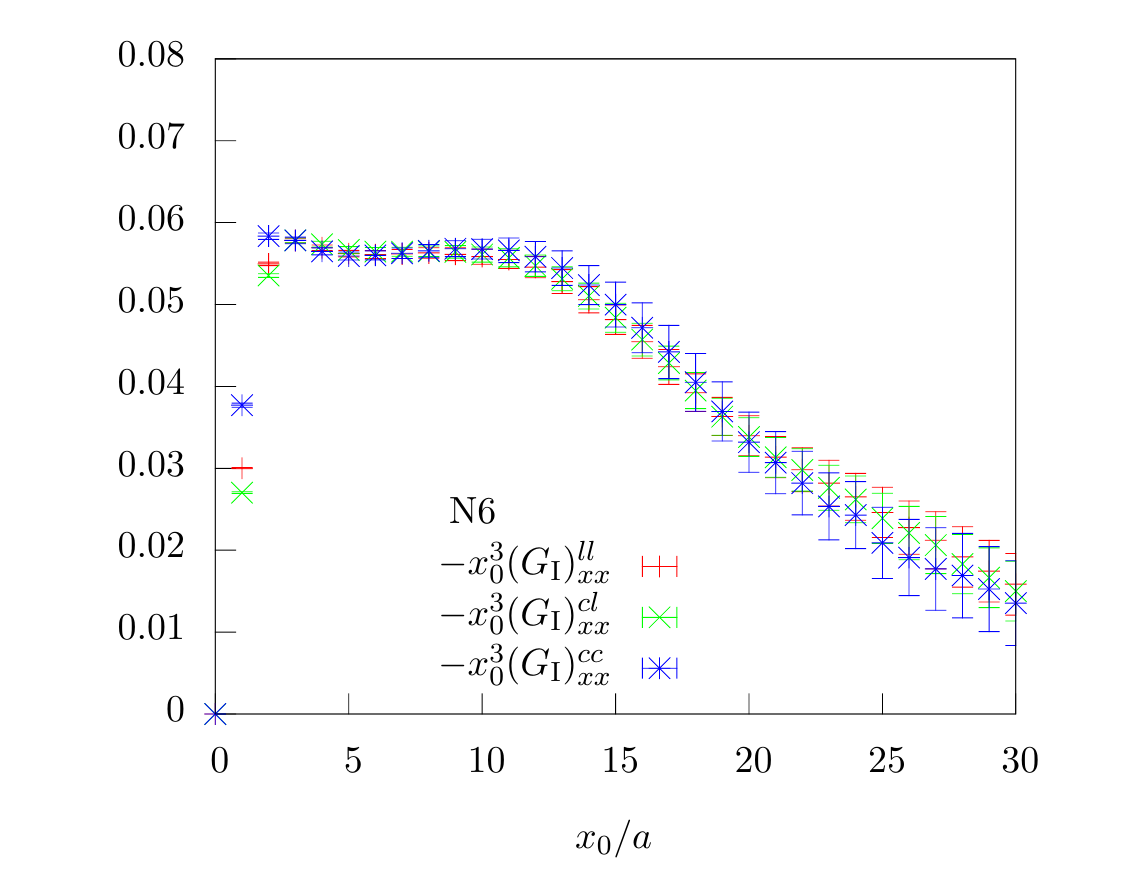}
    }
    \caption{Improved correlators A5 (left) and N6 (right).}
    \label{fig:improved_correlators}
\end{figure*}

In figure~\ref{fig:improved_correlators} we show the effect of the
improvement on the vector correlators with the given improvement
condition for the A5 (left) and N6 (right) ensembles.  By definition,
the central values now coincide at $x_0/a=3$.
The effect of the improvement appears to be smallest for the local-conserved vector
current correlator, due to the contributions of the improvement of
each current entering with opposite signs.

\subsubsection{Interpolation in $g_0^2$}
The following polynomial interpolation formulas in $g_0^2=6/\beta$ can be used
to determine the improvement coefficients for $N_\mathrm{f}=2$ flavours of
non-perturbatively improved Wilson fermions
\begin{align}
    \label{eq:interpolation}
    \cv{l} (g_0^2) &= -0.01225 \, C_\mathrm{F} g_0^2 \left( 1 + 7.19\,g_0^2 + 10.15\,g_0^4 \right),\\
    \cv{c} (g_0^2) &= \frac{1}{2} \left( 1 + 0.33\,g_0^2 -0.728\,g_0^4  \right).
\end{align}
Conservatively, these parametrizations should be used in the interval $5.2\leq \beta\leq 5.5$, even though the known behavior at small $g_0^2$ is built in.
It would be interesting to extend the present calculations to smaller values of $g_0^2$ to make explicit contact with the one-loop result.

\section{\label{sec:continuum_limit}Continuum limit of improved observables}

In order to quantify the effect of the improvement we examine the scaling of the observable
\begin{align}
    I_\mathrm{I}^{ij} = \int^{t_a}_{t_b} \mathrm{d}x_0 \, x_0^4 (G_\mathrm{I})^{ij}_{xx}(x_0)
    \label{eq:integral}
\end{align}
and its unimproved counterpart $I_\mathrm{U}^{ij}$, which is defined
analogously with the unimproved two-point function $(G_\mathrm{U})^{ij}_{xx}(x_0)$ (eq.~(\ref{eq:unimp_vector_correlator})),
toward the continuum limit.  The limits $t_a/a=8$ and $t_b/a=26$ are
fixed at the smallest lattice spacing, corresponding to $t_a\approx
0.4\mathrm{fm}$ and $t_b\approx1.3\mathrm{fm}$.  Although the contact
term does not contribute to such an observable when $t_a=0$, the lower limit
explicitly removes very short-distance contributions.  A lattice
estimate for this observable is obtained from the discretized correlation function by quadrature with an improved integration scheme.
An interpolation is needed at the limits for the coarser two lattice
spacings.  This observable is related to the slope of the Adler
function through the mixed representation of the hadronic vacuum
polarization function~\cite{Francis:2013fzp}.  Therefore, it may serve as a useful proxy to
quantify the effect of the improvement on phenomenologically relevant
observables which are dominated by the long-distance physics of the
vector correlation function.

The values of the Sommer scale, $r_0/a$, used to set the relative
scale and perform the continuum limit were taken from
ref.~\cite{Fritzsch:2012wq}.

\begin{figure*}[t]
    \centerline{
    \includegraphics[scale=0.8]{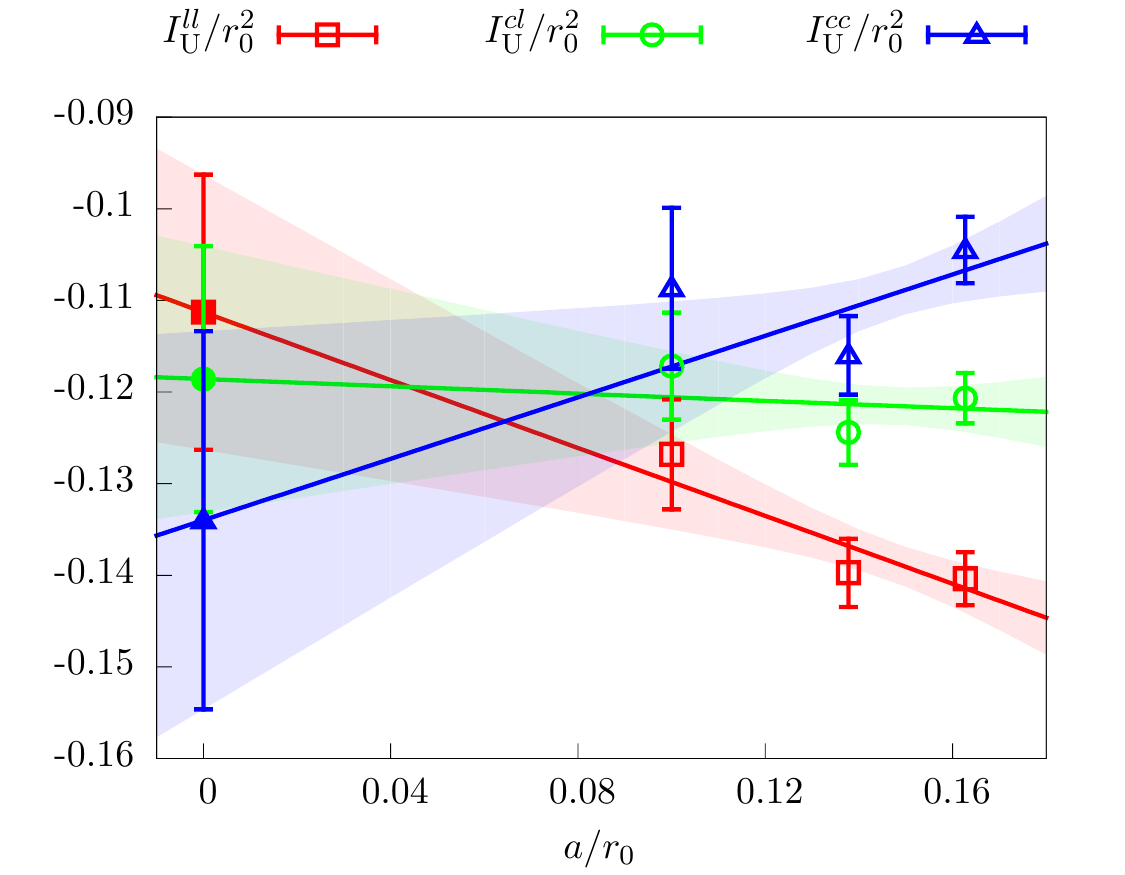}%
    \includegraphics[scale=0.8]{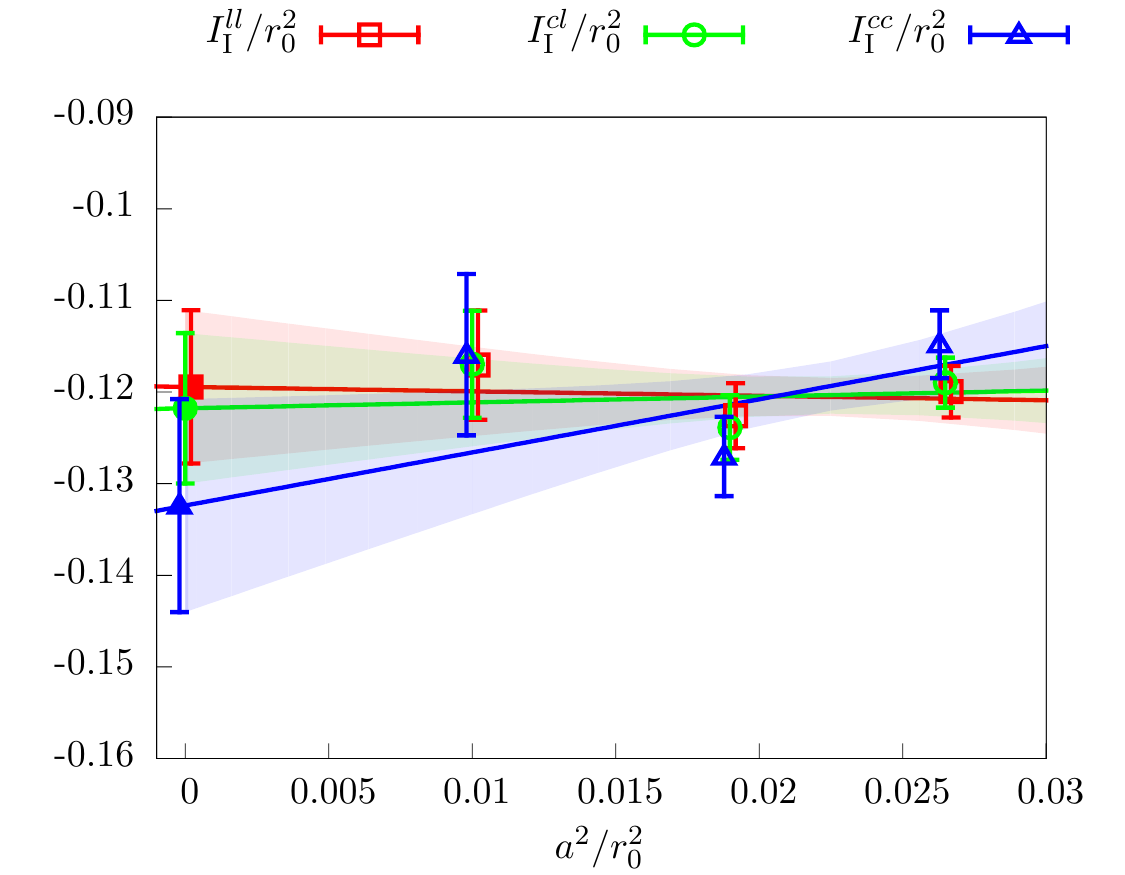}
    }
    \caption{Linear scaling of observable~(\ref{eq:integral}) for unimproved currents (left) and quadratic scaling of improved currents (right).
    The points in the right-hand panel have been displaced horizontally for clarity.}
    \label{fig:t4kernel}
\end{figure*}

\begin{figure*}[t]
    \centering
    \includegraphics[scale=0.8]{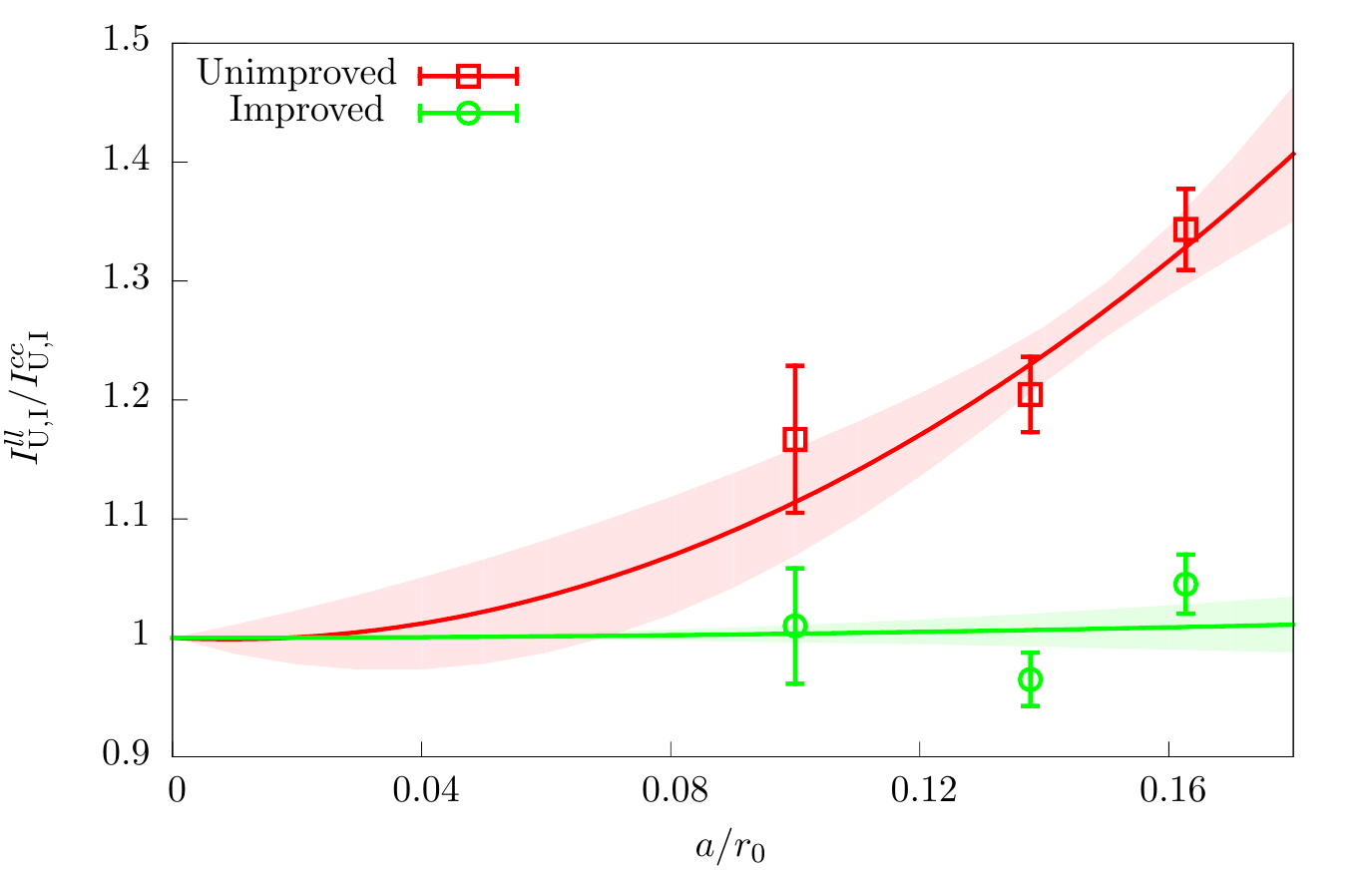}
    \caption{Scaling of the ratio of local-local and conserved-conserved observable~(\ref{eq:integral}).}
    \label{fig:ratio}
\end{figure*}

In figure~\ref{fig:t4kernel} (left) the three discretizations
$I_{\mathrm{U}}^{ll}$, $I_{\mathrm{U}}^{cl}$, $I_{\mathrm{U}}^{cc}$
are shown in red, green and blue, respectively.  The scaling of the
unimproved observables is modelled linearly in $a$ to obtain the
continuum limit.  While the continuum limits of the different
discretizations agree within the statistical precision, and could be
constrained to agree in a simultaneous fit, the use of a single
discretization demonstrates a significant fraction of the uncertainty
in the continuum result is due to the long extrapolation in $a$.

In the right-hand panel of figure~\ref{fig:t4kernel} the analogous
plot for the improved observable is shown.  A quadratic model in $a$
describes the continuum scaling well, and the error in continuum limit
is correspondingly reduced.  Furthermore, the residual scaling
violations appear to be small.  Note that the observable defined from
the local-conserved vector correlation function appears to have the
mildest scaling of all three unimproved discretizations.

Another illustration of the improved scaling behaviour of the
$\order{a}$-improved observable is shown in figure~\ref{fig:ratio}.
The ratio of the observable defined
using the local-local vector correlation function and the
conserved-conserved vector correlation function is shown for the
unimproved (red) and improved currents (green).  This discrepancy
should vanish in the continuum limit and the fits (solid lines) are
constrained to vanish there likewise.  Here, owing to the additional
constraint imposed in the continuum limit we can use a quadratic model
for both unimproved and improved discretizations.

\section{\label{sec:conclusions}Conclusions}

The basic idea of the improved strategy used here is that if $n_d$
discretizations of a current are considered, there are $n_d(n_d+1)/2$
lattice versions of its two-point function. Requiring the equality of
these two-point functions thus allows one to determine the improvement
coefficients of each discretization, for $n_d$ sufficiently large.  The
method however requires that the relative normalization of the
operators be known at the outset. For the program to go through
consistently one thus needs to determine the renormalization factors
in a situation where the improvement terms do not contribute.
This is the case for the vector current, because the improvement 
term does not affect the conserved charge.

Since the improvement of spectral quantities only depends on the
improvement of the action, the effect of the vector current
improvement is expected to be less significant for the effective mass
at moderate distances, where the vector correlator is dominated by the
rho meson.  This has implications for e.g. the scale-setting
procedure defined through the effective mass of the vector
correlator~\cite{Francis:2013jfa}.  We explicitly observed the effect
of the improvement in this quantity to be small.



\acknowledgments{We thank H.\ Wittig, D.\ Djukanovic, J.\ Green and all our colleagues in the Mainz lattice group
for helpful discussions, encouragement and support.
The correlation functions were computed on the ``Clover'' platform at the Helmholtz-Institute Mainz.
The work of H.M. is partly supported 
by the DFG under grants ME 3622/2-1 and ME 3622/2-2.}

\bibliography{cv}

\begin{thebibliography}{16}%
\makeatletter
\providecommand \@ifxundefined [1]{%
 \@ifx{#1\undefined}
}%
\providecommand \@ifnum [1]{%
 \ifnum #1\expandafter \@firstoftwo
 \else \expandafter \@secondoftwo
 \fi
}%
\providecommand \@ifx [1]{%
 \ifx #1\expandafter \@firstoftwo
 \else \expandafter \@secondoftwo
 \fi
}%
\providecommand \natexlab [1]{#1}%
\providecommand \enquote  [1]{``#1''}%
\providecommand \bibnamefont  [1]{#1}%
\providecommand \bibfnamefont [1]{#1}%
\providecommand \citenamefont [1]{#1}%
\providecommand \href@noop [0]{\@secondoftwo}%
\providecommand \href [0]{\begingroup \@sanitize@url \@href}%
\providecommand \@href[1]{\@@startlink{#1}\@@href}%
\providecommand \@@href[1]{\endgroup#1\@@endlink}%
\providecommand \@sanitize@url [0]{\catcode `\\12\catcode `\$12\catcode
  `\&12\catcode `\#12\catcode `\^12\catcode `\_12\catcode `\%12\relax}%
\providecommand \@@startlink[1]{}%
\providecommand \@@endlink[0]{}%
\providecommand \url  [0]{\begingroup\@sanitize@url \@url }%
\providecommand \@url [1]{\endgroup\@href {#1}{\urlprefix }}%
\providecommand \urlprefix  [0]{URL }%
\providecommand \Eprint [0]{\href }%
\providecommand \doibase [0]{http://dx.doi.org/}%
\providecommand \selectlanguage [0]{\@gobble}%
\providecommand \bibinfo  [0]{\@secondoftwo}%
\providecommand \bibfield  [0]{\@secondoftwo}%
\providecommand \translation [1]{[#1]}%
\providecommand \BibitemOpen [0]{}%
\providecommand \bibitemStop [0]{}%
\providecommand \bibitemNoStop [0]{.\EOS\space}%
\providecommand \EOS [0]{\spacefactor3000\relax}%
\providecommand \BibitemShut  [1]{\csname bibitem#1\endcsname}%
\let\auto@bib@innerbib\@empty
\bibitem [{\citenamefont {Symanzik}(1983{\natexlab{a}})}]{Symanzik:1983dc}%
  \BibitemOpen
  \bibfield  {author} {\bibinfo {author} {\bibfnamefont {K.}~\bibnamefont
  {Symanzik}},\ }\href {\doibase 10.1016/0550-3213(83)90468-6} {\bibfield
  {journal} {\bibinfo  {journal} {Nucl.Phys.}\ }\textbf {\bibinfo {volume}
  {B226}},\ \bibinfo {pages} {187} (\bibinfo {year}
  {1983}{\natexlab{a}})}\BibitemShut {NoStop}%
\bibitem [{\citenamefont {Symanzik}(1983{\natexlab{b}})}]{Symanzik:1983gh}%
  \BibitemOpen
  \bibfield  {author} {\bibinfo {author} {\bibfnamefont {K.}~\bibnamefont
  {Symanzik}},\ }\href {\doibase 10.1016/0550-3213(83)90469-8} {\bibfield
  {journal} {\bibinfo  {journal} {Nucl.Phys.}\ }\textbf {\bibinfo {volume}
  {B226}},\ \bibinfo {pages} {205} (\bibinfo {year}
  {1983}{\natexlab{b}})}\BibitemShut {NoStop}%
\bibitem [{\citenamefont {Wilson}(1974)}]{Wilson:1974sk}%
  \BibitemOpen
  \bibfield  {author} {\bibinfo {author} {\bibfnamefont {K.~G.}\ \bibnamefont
  {Wilson}},\ }\href {\doibase 10.1103/PhysRevD.10.2445} {\bibfield  {journal}
  {\bibinfo  {journal} {Phys.Rev.}\ }\textbf {\bibinfo {volume} {D10}},\
  \bibinfo {pages} {2445} (\bibinfo {year} {1974})}\BibitemShut {NoStop}%
\bibitem [{\citenamefont {Sheikholeslami}\ and\ \citenamefont
  {Wohlert}(1985)}]{Sheikholeslami:1985ij}%
  \BibitemOpen
  \bibfield  {author} {\bibinfo {author} {\bibfnamefont {B.}~\bibnamefont
  {Sheikholeslami}}\ and\ \bibinfo {author} {\bibfnamefont {R.}~\bibnamefont
  {Wohlert}},\ }\href {\doibase 10.1016/0550-3213(85)90002-1} {\bibfield
  {journal} {\bibinfo  {journal} {Nucl.Phys.}\ }\textbf {\bibinfo {volume}
  {B259}},\ \bibinfo {pages} {572} (\bibinfo {year} {1985})}\BibitemShut
  {NoStop}%
\bibitem [{\citenamefont {Luscher}\ \emph {et~al.}(1996)\citenamefont
  {Luscher}, \citenamefont {Sint}, \citenamefont {Sommer},\ and\ \citenamefont
  {Weisz}}]{Luscher:1996sc}%
  \BibitemOpen
  \bibfield  {author} {\bibinfo {author} {\bibfnamefont {M.}~\bibnamefont
  {Luscher}}, \bibinfo {author} {\bibfnamefont {S.}~\bibnamefont {Sint}},
  \bibinfo {author} {\bibfnamefont {R.}~\bibnamefont {Sommer}}, \ and\ \bibinfo
  {author} {\bibfnamefont {P.}~\bibnamefont {Weisz}},\ }\href {\doibase
  10.1016/0550-3213(96)00378-1} {\bibfield  {journal} {\bibinfo  {journal}
  {Nucl.Phys.}\ }\textbf {\bibinfo {volume} {B478}},\ \bibinfo {pages} {365}
  (\bibinfo {year} {1996})},\ \Eprint {http://arxiv.org/abs/hep-lat/9605038}
  {arXiv:hep-lat/9605038 [hep-lat]} \BibitemShut {NoStop}%
\bibitem [{\citenamefont {Guagnelli}\ and\ \citenamefont
  {Sommer}(1998)}]{Guagnelli:1997db}%
  \BibitemOpen
  \bibfield  {author} {\bibinfo {author} {\bibfnamefont {M.}~\bibnamefont
  {Guagnelli}}\ and\ \bibinfo {author} {\bibfnamefont {R.}~\bibnamefont
  {Sommer}},\ }\href {\doibase 10.1016/S0920-5632(97)00930-4} {\bibfield
  {journal} {\bibinfo  {journal} {Nucl.Phys.Proc.Suppl.}\ }\textbf {\bibinfo
  {volume} {63}},\ \bibinfo {pages} {886} (\bibinfo {year} {1998})},\ \Eprint
  {http://arxiv.org/abs/hep-lat/9709088} {arXiv:hep-lat/9709088 [hep-lat]}
  \BibitemShut {NoStop}%
\bibitem [{\citenamefont {Bhattacharya}\ \emph {et~al.}(1999)\citenamefont
  {Bhattacharya}, \citenamefont {Chandrasekharan}, \citenamefont {Gupta},
  \citenamefont {Lee},\ and\ \citenamefont {Sharpe}}]{Bhattacharya:1999uq}%
  \BibitemOpen
  \bibfield  {author} {\bibinfo {author} {\bibfnamefont {T.}~\bibnamefont
  {Bhattacharya}}, \bibinfo {author} {\bibfnamefont {S.}~\bibnamefont
  {Chandrasekharan}}, \bibinfo {author} {\bibfnamefont {R.}~\bibnamefont
  {Gupta}}, \bibinfo {author} {\bibfnamefont {W.-J.}\ \bibnamefont {Lee}}, \
  and\ \bibinfo {author} {\bibfnamefont {S.~R.}\ \bibnamefont {Sharpe}},\
  }\href {\doibase 10.1016/S0370-2693(99)00796-0} {\bibfield  {journal}
  {\bibinfo  {journal} {Phys.Lett.}\ }\textbf {\bibinfo {volume} {B461}},\
  \bibinfo {pages} {79} (\bibinfo {year} {1999})},\ \Eprint
  {http://arxiv.org/abs/hep-lat/9904011} {arXiv:hep-lat/9904011 [hep-lat]}
  \BibitemShut {NoStop}%
\bibitem [{\citenamefont {Jansen}\ and\ \citenamefont
  {Sommer}(1998)}]{Jansen:1998mx}%
  \BibitemOpen
  \bibfield  {author} {\bibinfo {author} {\bibfnamefont {K.}~\bibnamefont
  {Jansen}}\ and\ \bibinfo {author} {\bibfnamefont {R.}~\bibnamefont {Sommer}}
  (\bibinfo {collaboration} {ALPHA}),\ }\href {\doibase
  10.1016/S0550-3213(98)00396-4} {\bibfield  {journal} {\bibinfo  {journal}
  {Nucl.Phys.}\ }\textbf {\bibinfo {volume} {B530}},\ \bibinfo {pages} {185}
  (\bibinfo {year} {1998})},\ \Eprint {http://arxiv.org/abs/hep-lat/9803017}
  {arXiv:hep-lat/9803017 [hep-lat]} \BibitemShut {NoStop}%
\bibitem [{\citenamefont {Sint}\ and\ \citenamefont
  {Weisz}(1997)}]{Sint:1997jx}%
  \BibitemOpen
  \bibfield  {author} {\bibinfo {author} {\bibfnamefont {S.}~\bibnamefont
  {Sint}}\ and\ \bibinfo {author} {\bibfnamefont {P.}~\bibnamefont {Weisz}},\
  }\href {\doibase 10.1016/S0550-3213(97)00372-6} {\bibfield  {journal}
  {\bibinfo  {journal} {Nucl.Phys.}\ }\textbf {\bibinfo {volume} {B502}},\
  \bibinfo {pages} {251} (\bibinfo {year} {1997})},\ \Eprint
  {http://arxiv.org/abs/hep-lat/9704001} {arXiv:hep-lat/9704001 [hep-lat]}
  \BibitemShut {NoStop}%
\bibitem [{\citenamefont {Francis}\ \emph
  {et~al.}(2013{\natexlab{a}})\citenamefont {Francis}, \citenamefont {von
  Hippel}, \citenamefont {Meyer},\ and\ \citenamefont
  {Jegerlehner}}]{Francis:2013jfa}%
  \BibitemOpen
  \bibfield  {author} {\bibinfo {author} {\bibfnamefont {A.}~\bibnamefont
  {Francis}}, \bibinfo {author} {\bibfnamefont {G.}~\bibnamefont {von Hippel}},
  \bibinfo {author} {\bibfnamefont {H.~B.}\ \bibnamefont {Meyer}}, \ and\
  \bibinfo {author} {\bibfnamefont {F.}~\bibnamefont {Jegerlehner}},\
  }\href@noop {} {\bibfield  {journal} {\bibinfo  {journal} {PoS}\ }\textbf
  {\bibinfo {volume} {LATTICE2013}},\ \bibinfo {pages} {320} (\bibinfo {year}
  {2013}{\natexlab{a}})},\ \Eprint {http://arxiv.org/abs/1312.0035}
  {arXiv:1312.0035 [hep-lat]} \BibitemShut {NoStop}%
\bibitem [{\citenamefont {Della~Morte}\ \emph {et~al.}(2005)\citenamefont
  {Della~Morte}, \citenamefont {Hoffmann}, \citenamefont {Knechtli},
  \citenamefont {Sommer},\ and\ \citenamefont {Wolff}}]{DellaMorte:2005rd}%
  \BibitemOpen
  \bibfield  {author} {\bibinfo {author} {\bibfnamefont {M.}~\bibnamefont
  {Della~Morte}}, \bibinfo {author} {\bibfnamefont {R.}~\bibnamefont
  {Hoffmann}}, \bibinfo {author} {\bibfnamefont {F.}~\bibnamefont {Knechtli}},
  \bibinfo {author} {\bibfnamefont {R.}~\bibnamefont {Sommer}}, \ and\ \bibinfo
  {author} {\bibfnamefont {U.}~\bibnamefont {Wolff}},\ }\href {\doibase
  10.1088/1126-6708/2005/07/007} {\bibfield  {journal} {\bibinfo  {journal}
  {JHEP}\ }\textbf {\bibinfo {volume} {0507}},\ \bibinfo {pages} {007}
  (\bibinfo {year} {2005})},\ \Eprint {http://arxiv.org/abs/hep-lat/0505026}
  {arXiv:hep-lat/0505026 [hep-lat]} \BibitemShut {NoStop}%
\bibitem [{\citenamefont {{Dalla Brida}}\ and\ \citenamefont
  {Sint}(2014)}]{Brida:2014zwa}%
  \BibitemOpen
  \bibfield  {author} {\bibinfo {author} {\bibfnamefont {M.}~\bibnamefont
  {{Dalla Brida}}}\ and\ \bibinfo {author} {\bibfnamefont {S.}~\bibnamefont
  {Sint}},\ }\href@noop {} {\bibfield  {journal} {\bibinfo  {journal} {PoS}\
  }\textbf {\bibinfo {volume} {LATTICE2014}},\ \bibinfo {pages} {280} (\bibinfo
  {year} {2014})},\ \Eprint {http://arxiv.org/abs/1412.8022} {arXiv:1412.8022
  [hep-lat]} \BibitemShut {NoStop}%
\bibitem [{\citenamefont {Gockeler}\ \emph {et~al.}(2004)\citenamefont
  {Gockeler} \emph {et~al.}}]{Gockeler:2003cw}%
  \BibitemOpen
  \bibfield  {author} {\bibinfo {author} {\bibfnamefont {M.}~\bibnamefont
  {Gockeler}} \emph {et~al.} (\bibinfo {collaboration} {QCDSF}),\ }\href
  {\doibase 10.1016/j.nuclphysb.2004.03.026} {\bibfield  {journal} {\bibinfo
  {journal} {Nucl.Phys.}\ }\textbf {\bibinfo {volume} {B688}},\ \bibinfo
  {pages} {135} (\bibinfo {year} {2004})},\ \Eprint
  {http://arxiv.org/abs/hep-lat/0312032} {arXiv:hep-lat/0312032 [hep-lat]}
  \BibitemShut {NoStop}%
\bibitem [{Note1()}]{Note1}%
  \BibitemOpen
  \bibinfo {note} {It would however not be possible to use a point-split tensor
  current for the improved operator in conjunction with the improvement
  coefficient determined from the local one. These definitions differ already
  at $\protect \mathrm {O}(g_0^2)$.}\BibitemShut {Stop}%
\bibitem [{\citenamefont {Fritzsch}\ \emph {et~al.}(2012)\citenamefont
  {Fritzsch}, \citenamefont {Knechtli}, \citenamefont {Leder}, \citenamefont
  {Marinkovic}, \citenamefont {Schaefer} \emph {et~al.}}]{Fritzsch:2012wq}%
  \BibitemOpen
  \bibfield  {author} {\bibinfo {author} {\bibfnamefont {P.}~\bibnamefont
  {Fritzsch}}, \bibinfo {author} {\bibfnamefont {F.}~\bibnamefont {Knechtli}},
  \bibinfo {author} {\bibfnamefont {B.}~\bibnamefont {Leder}}, \bibinfo
  {author} {\bibfnamefont {M.}~\bibnamefont {Marinkovic}}, \bibinfo {author}
  {\bibfnamefont {S.}~\bibnamefont {Schaefer}},  \emph {et~al.},\ }\href
  {\doibase 10.1016/j.nuclphysb.2012.07.026} {\bibfield  {journal} {\bibinfo
  {journal} {Nucl.Phys.}\ }\textbf {\bibinfo {volume} {B865}},\ \bibinfo
  {pages} {397} (\bibinfo {year} {2012})},\ \Eprint
  {http://arxiv.org/abs/1205.5380} {arXiv:1205.5380 [hep-lat]} \BibitemShut
  {NoStop}%
\bibitem [{\citenamefont {Francis}\ \emph
  {et~al.}(2013{\natexlab{b}})\citenamefont {Francis}, \citenamefont {Jaeger},
  \citenamefont {Meyer},\ and\ \citenamefont {Wittig}}]{Francis:2013fzp}%
  \BibitemOpen
  \bibfield  {author} {\bibinfo {author} {\bibfnamefont {A.}~\bibnamefont
  {Francis}}, \bibinfo {author} {\bibfnamefont {B.}~\bibnamefont {Jaeger}},
  \bibinfo {author} {\bibfnamefont {H.~B.}\ \bibnamefont {Meyer}}, \ and\
  \bibinfo {author} {\bibfnamefont {H.}~\bibnamefont {Wittig}},\ }\href
  {\doibase 10.1103/PhysRevD.88.054502} {\bibfield  {journal} {\bibinfo
  {journal} {Phys.Rev.}\ }\textbf {\bibinfo {volume} {D88}},\ \bibinfo {pages}
  {054502} (\bibinfo {year} {2013}{\natexlab{b}})},\ \Eprint
  {http://arxiv.org/abs/1306.2532} {arXiv:1306.2532 [hep-lat]} \BibitemShut
  {NoStop}%
\end{thebibliography}%

\end{document}